\documentclass[aip,apl,reprint]{revtex4-1} 

\usepackage{graphicx}

\begin{document}

\title{Ultrafast inhomogeneous magnetization dynamics analyzed by interface-sensitive nonlinear magneto-optics} 

\author{J. Chen}
\affiliation{Faculty of Physics and Center for Nanointegration
(CENIDE), University of Duisburg-Essen, Lotharstr.~1, 47057
Duisburg, Germany}

\author{J. Wieczorek}
\affiliation{Faculty of Physics and Center for Nanointegration
(CENIDE), University of Duisburg-Essen, Lotharstr.~1, 47057
Duisburg, Germany}

\author{A. Eschenlohr} 
\email[]{andrea.eschenlohr@uni-due.de}
\homepage[]{www.uni-due.de/agbovensiepen}
\affiliation{Faculty of Physics and Center for Nanointegration
(CENIDE), University of Duisburg-Essen, Lotharstr.~1, 47057
Duisburg, Germany}

\author{S. Xiao}
\affiliation{Faculty of Physics and Center for Nanointegration
(CENIDE), University of Duisburg-Essen, Lotharstr.~1, 47057
Duisburg, Germany}

\author{A. Tarasevitch}
\affiliation{Faculty of Physics and Center for Nanointegration
(CENIDE), University of Duisburg-Essen, Lotharstr.~1, 47057
Duisburg, Germany}

\author{U. Bovensiepen} 
\affiliation{Faculty of Physics and Center for
Nanointegration (CENIDE), University of Duisburg-Essen,
Lotharstr.~1, 47057 Duisburg, Germany}

\date{\today}

\begin{abstract}
We analyze laser-induced ultrafast, spatially inhomogeneous magnetization dynamics of epitaxial Co/Cu(001) films in a 0.4-10~nm thickness range with time-resolved magnetization-induced second harmonic generation, which probes femtosecond spin dynamics at the vacuum/Co and Co/Cu interfaces. The interference of these two contributions makes the overall signal particularly sensitive to differences in the transient magnetization redistribution between the two interfaces, i.e. ultrafast magnetization profiles in the ferromagnetic film. We find in films of up to 3~nm thickness a stronger demagnetization at the surface, because the film thickness is smaller than the effective mean free path of the spin current mediating the demagnetization, i.e. the difference between the mean free paths of the majority and minority carriers. For film thicknesses larger than 3~nm, the magnetization profile reverses, since majority spins can escape into the conducting substrate only from the interface-near region. 
\end{abstract}

\pacs{72.25.Ba, 72.25.Fe, 72.25.Mk, 75.78.Jp, 78.20.Ls} 

\maketitle 


The observation of optically excited ultrafast spin currents in ferromagnetic (FM) metals \cite{battiato2010, melnikov2011} has opened up new possibilities to manipulate magnetization on sub-picosecond timescales. Both transient magnetization enhancement \cite{rudolf2012} and ultrafast demagnetization \cite{malinowski2008, eschenlohr2013} by such spin-dependent, screened \cite{battiato2012} charge currents, as well as spin-transfer torque \cite{schellekens2014} in metallic layered structures, were recently demonstrated. These discoveries make it conceivable to extend spintronics into the ultrafast regime, employing highly excited, non-equilibrium spin-polarized carriers. In this context, analyzing spin dynamics at interfaces has become increasingly important, as the interface properties are decisive for spin transport in heterostructures potentially relevant for applications, e.g. employing spin valves or multilayers \cite{belien1994, zahn1998}. An experimental technique capable of addressing this issue must be interface sensitive, as well as able to probe spin dynamics at buried interfaces, on the relevant fs timescales in a pump-probe experiment. \\
Here, we show that magnetization-induced second harmonic generation (mSHG) \cite{pan1989} can probe and even distinguish contributions to spin dynamics from several interfaces in a ferromagnet-metal heterostructure by performing a fs time-resolved, thickness-dependent study on epitaxial Co/Cu(001) films as a model system. Based on this analysis, we identify transient spatial magnetization profiles in the direction normal to the sample surface during ultrafast demagnetization. We find that these profiles reflect the effective escape depth of fs spin currents into Cu via the difference in the spin-dependent mean free paths (MFP) of majority and minority electrons, thus changing strongly with increasing sample thickness. \\
\begin{figure}
	\includegraphics{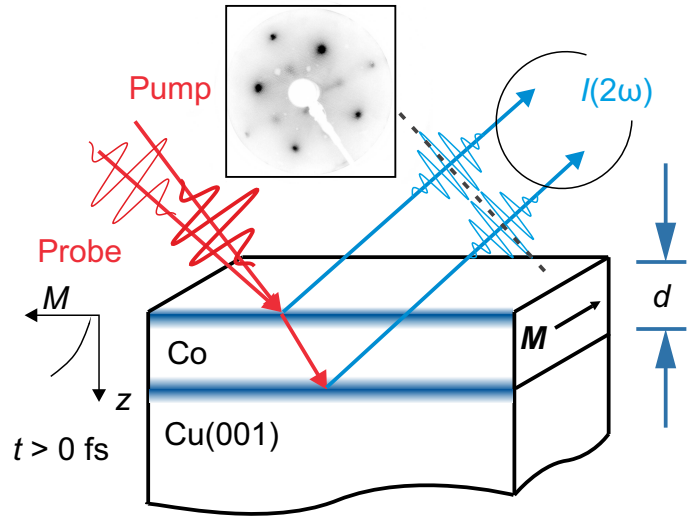}
	\caption{\label{fig1} Schematic experimental setup for measuring the interface-sensitive mSHG yield in reflection from Co/Cu(001) for different Co film thicknesses $d$ ranging from 0.4~nm to 10~nm in a pump-probe experiment. Ultrafast demagnetization due to the pump pulse will lead to a spatially inhomogeneous change in the magnetization $M$ along the film depth $z$ \cite{wieczorek2015}. (inset) LEED pattern at 128~eV kinectic energy of 0.9~nm thick Co/Cu(001), showing the four-fold symmetry expected from epitaxial growth of Co.}
\end{figure}
The schematic experimental arrangement is depicted in Fig. \ref{fig1}. Ultrashort laser pulses were generated from a cavity-dumped Ti:Sa oscillator, with a pulse width of 35~fs at a central wavelength of 800~nm, 40 nJ pulse energy and a repetition rate of 2.53~MHz. The pump and probe beams were split at a 4:1 intensity ratio. The incident pump fluence at the sample surface was 6 mJ/cm$^{2}$, and the pump pulses were s-polarized. We measured the reflected second harmonic (SH) yield at 400~nm wavelength from the sample, using a BG39 optical filter and a grating monochromator, by single photon counting. The measurements were performed in the transversal geometry, i.e. with the magnetization $M$ aligned in the sample plane and normal to the optical plane of incidence by an applied magnetic field large enough to reverse $M$. The ingoing probe pulse was p-polarized, and p-polarized SH radiation was detected in the static experiments. In the pump-probe experiments, detection was performed without polarization analysis. Co films with thicknesses $0.4$~nm $\leq d \leq 10$~nm were grown \textit{in situ} by electron beam evaporation in ultrahigh vacuum ($<10^{-10}$~mbar) on a Cu(001) single crystal prepared by sputtering-annealing cycles \cite{gonzalez1992}. Epitaxial growth was verified by low energy electron diffraction (LEED), see Fig. \ref{fig1}. \\ 
Essential for the interpretation of our time-resolved mSHG results is the fact that the overall detected SH radiation is generated at two interfaces, namely vacuum/Co and Co/Cu, in our Co/Cu(001) samples, see Fig. \ref{fig1}, and that the respective magnetization-dependent SH fields interfere with each other. We will first discuss the static thickness dependence, which allows us to describe this interference. \\ 
The magnetization-dependent SH field $\left|E_{\mathrm{odd}}\right| \cdot \mathrm{cos}(\alpha) \approx \frac{I^{\uparrow} - I^{\downarrow}}{4\left|E_{\mathrm{even}}\right|}$, with the magnetization-\textit{in}dependent SH field $\left|E_{\mathrm{even}}\right| \approx \sqrt{\frac{I^{\uparrow} + I^{\downarrow}}{2}}$, is determined from the measured SH intensities $I^{\uparrow,\downarrow}$ at opposite orientations of $M$. Its thickness dependence for $0.4$~nm $\leq d \leq 10$~nm is displayed in Fig. \ref{fig2}. Here, $\alpha$ refers to the phase between $\left|E_{\mathrm{even}}\right|$ and $\left|E_{\mathrm{odd}}\right|$, which is close to zero and constant for small $d$ \cite{conrad2001}. We thus assume $\alpha = 0$ in the following. $\left|E_{\mathrm{odd}}\right|$ decreases sharply until about $d = 3$~nm, then increases before staying at nearly constant values for larger $d$. In order to explain this behavior, we will describe $E_{\mathrm{odd}}$ in terms of the contributions from the Co surface $E^{\mathrm{S}}_{\mathrm{odd}}$ and the Co/Cu interface $E^{\mathrm{I}}_{\mathrm{odd}}$, i.e. $E_{\mathrm{odd}} = E^{\mathrm{S}}_{\mathrm{odd}} + E^{\mathrm{I}}_{\mathrm{odd}}$, where $E^{\mathrm{S}}_{\mathrm{odd}} = a_{\mathrm{S}} \cdot m_{\mathrm{S}} \cdot E^{2}_{\mathrm{\omega, S}}$ and $E^{\mathrm{I}}_{\mathrm{odd}} = a_{\mathrm{I}} \cdot m_{\mathrm{I}} \cdot E^{2}_{\mathrm{\omega, I}} \cdot U(d)$, with $m_{\mathrm{S}}$ and $m_{\mathrm{I}}$ being the magnetization, and $E_{\mathrm{\omega, S}}$ and $E_{\mathrm{\omega, I}}$ the electric field of the fundamental pulse, at the surface respectively the interface. The proportionality factors $a_{\mathrm{S}}$ and $a_{\mathrm{I}}$ are given by the Fresnel factors and the symmetry-allowed magnetization-dependent elements of the second-order nonlinear susceptibility tensor \cite{huebner1994}. $U(d)$ describes the damping and phase shift of the SH field generated by the Co film. Consequently, mSHG preferentially probes the surface contribution with increasing $d$, because the SH from the interface is increasingly damped. Moreover, additional effects can occur due to interference of the surface and phase-shifted interface contributions. \\ 
We combine the above to an effective description with  
\begin{equation}
	\left|E_{\mathrm{odd}}(d)\right| = \left| A^{\mathrm{S}}_{\mathrm{odd}} + A^{\mathrm{I}}_{\mathrm{odd}} \cdot \mathrm{e}^{-1.29 \beta d} \cdot \mathrm{e}^{\mathrm{i} (0.017 \varphi + 0.077 d)} \right|, 
	\label{eq:fit} 
\end{equation}
where $A^{\mathrm{S}}_{\mathrm{odd}} = a_{\mathrm{S}} \cdot E^{2}_{\mathrm{\omega, S}}$, $A^{\mathrm{I}}_{\mathrm{odd}} = a_{\mathrm{I}} \cdot E^{2}_{\mathrm{\omega, I}}$, $\beta$ the effective damping in Co and $\varphi$ the effective relative phase shift between the surface and interface contributions, and fit our static mSHG data, see Fig. \ref{fig2}. The resulting value is $\varphi = 186\pm8^{\circ}$, showing a destructive interference of the surface and interface contributions. 
The value of $\varphi \approx 180^{\circ}$ occurs due to the incident beam being reflected at the FM layer at the vacuum/Co interface, and conversely at Cu after propagating through Co, in agreement with earlier findings \cite{wierenga1994}. \\
\begin{figure}
	\includegraphics{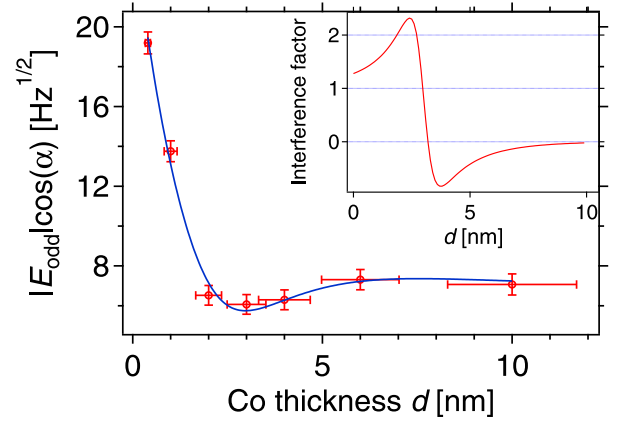}
	\caption{\label{fig2}Static measurement (circles) of the odd part of the SH field depending on the Co film thickness $d$. The solid line is a fit to the data according to equation \ref{eq:fit}. The inset depicts the calculated interference factor $int(d)$ between the odd contributions from the vacuum/Co and the Co/Cu interfaces.}
\end{figure}
Now, we turn our attention to the time-resolved mSHG response, shown in Fig. \ref{fig3} (top). $\Delta_{\mathrm{odd}}$ stands for the pump-induced relative change of $\left|E_{\mathrm{odd}}\right| \cdot cos(\alpha)$, normalized to its value before laser excitation. It is obvious that at various thicknesses, $\Delta_{\mathrm{odd}}$ shows distinctively different behavior. For $d=0.4$~nm, which equals two monolayers, a negative change of $\Delta_{\mathrm{odd}}$ occurs in the first few tens of fs after laser excitation, as expected for ultrafast demagnetization of the Co film, before a slower relaxation back to the equilibrium value. However, $\Delta_{\mathrm{odd}}$ for 2~nm $\leq d \leq$ 4~nm shows a positive change. For $d \geq 6$~nm, a sign change in the transient response occurs. We argue that this time- and $d$-dependent behavior is caused by spatially inhomogeneous magnetization dynamics in the FM film, which can in particular occur due to laser-induced spin transport and spin-flip scattering \cite{melnikov2011, battiato2012, wieczorek2015}. Since $E_{\mathrm{odd}}$ is very sensitive to differences between $m_{\mathrm{S}}$ and $m_{\mathrm{I}}$ due to the destructive interference of the SH fields from the two interfaces, our experiment is particularly sensitive to such effects. \\
To substantiate that the evolution of $\Delta_{\mathrm{odd}}$ is given by the transient relative changes of the magnetization at the interface $\Delta m_{\mathrm{I}} = (m_{\mathrm{I}}/m_{\mathrm{I}}(t < 0))-1$ and the surface $\Delta m_{\mathrm{S}} = (m_{\mathrm{S}}/m_{\mathrm{S}}(t < 0))-1 $, we approximate our time-dependent data as follows. As discussed above, the mSHG signal is dominated by the interface at low $d$ and the surface at higher $d$. Therefore, we take $\Delta_{\mathrm{odd}}$(10~nm) and $\Delta_{\mathrm{odd}}$(0.4~nm) as approximations of $\Delta^{\mathrm{S}}_{\mathrm{odd}}$ and $\Delta^{\mathrm{I}}_{\mathrm{odd}}$, respectively. It should thus be possible to describe the experimental data with a linear combination of the two terms, 
\begin{equation}
	\Delta_{\mathrm{odd}}(d)=a\Delta_{\mathrm{odd}}(10~\mathrm{nm})+b\Delta_{\mathrm{odd}}(0.4~\mathrm{nm}), 
	\label{eq:lincomb}
\end{equation}
where $a$ and $b$ represent the respective contributions from surface and interface, which change with thickness. As shown in Fig. \ref{fig3} (bottom) for intermediate $d$, the linear combination of $\Delta_{\mathrm{odd}}$(10~nm) and $\Delta_{\mathrm{odd}}$(0.4~nm) agrees well with the experimental data for $0.4$~nm $< d < 10$~nm at short delays up to about 0.5~ps. Only at longer delays, when the hot electron population thermalizes with the lattice, a deviation can be found. This agreement supports our assumption of interfering interface contributions also for dynamic, laser-induced changes, at least until lattice heating sets in. The thickness dependence of $\left|b\right|/(\left|a\right|+\left|b\right|)$, see Eq. \ref{eq:lincomb}, is displayed in Fig. \ref{fig4} (left). It clearly shows a continuous decrease of the interface contribution to $\Delta_{\mathrm{odd}}$ with increasing $d$, as expected. \\
In order to analyze the transient magnetization profiles \cite{melnikov2011, battiato2012, wieczorek2015}, we approximate $\Delta_{\mathrm{odd}}$ in terms of $\Delta m_{\mathrm{S}}$ and $\Delta m_{\mathrm{I}}$ in first order as 
\begin{equation}
	\Delta_{\mathrm{odd}}(t,d) \approx \Delta m_{\mathrm{S}}(t) + int(d) \cdot (\Delta m_{\mathrm{I}}(t) - \Delta m_{\mathrm{S}}(t)). 
	\label{eq:int}
\end{equation} 
We term $int(d)$ the interference factor, which determines how transient magnetization profiles, i.e. differences in $\Delta m_{\mathrm{S}}$ and $\Delta m_{\mathrm{I}}$, are expressed in $\Delta_{\mathrm{odd}}$ for a certain $d$. The interference factor describes the relative contribution of the damped and $180^{\circ}$ phase-shifted SH from the interface to the overall signal. It is calculated as $int(d) = \Re(B^{\mathrm{I}}_{\mathrm{odd}}(d)/(A^{\mathrm{S}}_{\mathrm{odd}} + B^{\mathrm{I}}_{\mathrm{odd}}(d))$ from the fit results according to Eq. \ref{eq:fit}, with $B^{\mathrm{I}}_{\mathrm{odd}}(d)$ referring to the second term on the right hand side of Eq. \ref{eq:fit},  and displayed in the inset of Fig. \ref{fig2}. The $d$-dependent changes in $\Delta_{\mathrm{odd}}(t)$ shown in Fig. \ref{fig3} (top) can now be understood with the aid of Eq. \ref{eq:int}: The expected negative change due to demagnetization can be more than compensated by a difference in the amount of demagnetization between surface and interface, which enters the observed signal via $int(d)$. In order to illustrate how mSHG probes magnetization profiles $\Delta m_{\mathrm{I}} - \Delta m_{\mathrm{S}}$, we rewrite equation \ref{eq:int} as 
\begin{equation}
	\Delta m_{\mathrm{I}}(t) - \Delta m_{\mathrm{S}}(t) \approx (\Delta_{\mathrm{odd}}(t) - \Delta m_{av}(t))	/ (int(d) - 0.5), 
	\label{eq:int2}
\end{equation} 
with $\Delta m_{av}$ referring to the average relative magnetization change in the film. We can thus calculate $\Delta m_{\mathrm{I}} - \Delta m_{\mathrm{S}}$, provided we have an observable for $\Delta m_{av}$. For $d\leq4$~nm, the transversal magneto-optical Kerr effect (T-MOKE) acquired simultaneously with the mSHG is shown in Fig. \ref{fig4}. It has a rather uniform depth sensitivity \cite{eschenlohr2016} and therefore serves as a good approximation for $\Delta m_{av}(t)$. \\ 
\begin{figure}
	\includegraphics{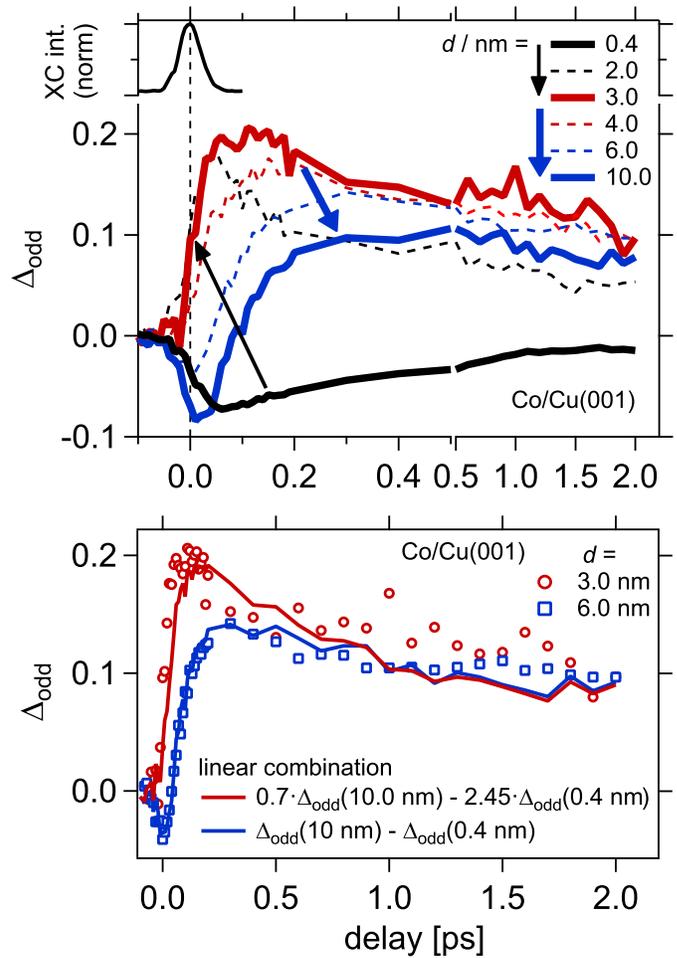}
	\caption{\label{fig3}(top) Time-resolved mSHG data $\Delta_{\mathrm{odd}}(t,d)$ versus the pump-probe time delay. The pump-probe cross-correlation (XC) indicates the experimental time resolution. Arrows indicate the evolution of the time-dependent $\Delta_{\mathrm{odd}}(t,d)$ signal with increasing thickness $d$. (bottom) Modelling of $\Delta_{\mathrm{odd}}(d)$ by linear combinations of $\Delta_{\mathrm{odd}}$(0.4~nm) and $\Delta_{\mathrm{odd}}$(10~nm) (lines) as described in the main text, in comparison to the experimental data (symbols).}
\end{figure}
From the results of the calculation according to equation \ref{eq:int2}, shown in Fig. \ref{fig4} (right) for 2 and 4~nm thickness, we see that for $d\leq3$~nm $\left|\Delta m_{\mathrm{S}}\right| > \left|\Delta m_{\mathrm{I}}\right|$, while for $d\geq4$~nm $\left|\Delta m_{\mathrm{S}}\right| < \left|\Delta m_{\mathrm{I}}\right|$ and thus $\Delta m_{\mathrm{I}} - \Delta m_{\mathrm{S}}$ reverses. We can conclude that the sign changes in $\Delta m_{\mathrm{I}} - \Delta m_{\mathrm{S}}$ are connected to the inelastic, spin-dependent electron MFP, if we consider that ultrafast demagnetization of thin ferromagnetic films on conducting substrates is driven primarily by fs spin currents before electron thermalization occurs \cite{battiato2010, melnikov2011, battiato2012, wieczorek2015}, i.e. majority spins escape into the substrate and cause a loss of spin polarization. For thicknesses which approach these MFPs of typically a few nm in the 3$d$ transition metal ferromagnets \cite{zhukov2006, kaltenborn2014}, it becomes less likely that majority spins propagate from the surface into the substrate before scattering. Instead a spatially inhomogeneous demagnetization which is stronger in the region near the interface to the substrate, i.e. the spin current sink, occurs. This effect can also be enhanced by the fact the minority spins with lower MFPs compared to the majority spins \cite{zhukov2006, kaltenborn2014} cannot leave the Co films for larger $d$ and accumulate near the buried interface \cite{eschenlohr2013}. Thus, the transient magnetization profiles are strongly influenced by the difference between the majority and minority electron MFPs. Our finding that the sign of $\Delta m_{\mathrm{I}} - \Delta m_{\mathrm{S}}$ reverses between $d=3$~nm and $d=4$~nm, compare Fig. \ref{fig4} (right), is consistent with the minority electron MFP being approximately 1-2~nm \cite{zhukov2006}, while the majority electron MFP is a factor of about 1.2 to 3 larger in the relevant energy range of up to 1.5~eV above the Fermi level \cite{zhukov2006, kaltenborn2014, goris2011}.\\
\begin{figure}
	\includegraphics{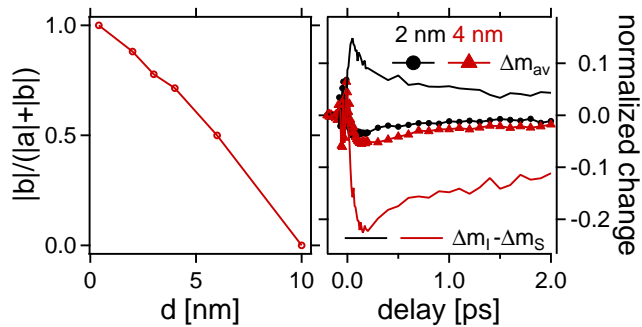}
	\caption{\label{fig4} (left) Ratio of the normalized linear combination factor $\left|b\right|$ over Co thickness. (right) $\Delta m_{av}$ for $d=2, 4$~nm Co/Cu(001) (symbols) together with $\Delta m_{\mathrm{I}} - \Delta m_{\mathrm{S}}$ calculated according to equation \ref{eq:int2} (lines).}
\end{figure}
For $d\geq6$~nm, our analysis according to equation \ref{eq:int2} is not suitable due to the fact that T-MOKE probes the demagnetization in an increasingly spatially inhogeneous manner and does not provide a satisfactory approximation for $\Delta m_{av}$ any more. However, the transient change shown for these $d$, compare Fig. \ref{fig3} (top), is still consistent with our previous results on Co/Cu(001) \cite{wieczorek2015}, which show a transiently changing spatial magnetization gradient due to the interplay between spin transport and phonon-mediated spin-flip scattering. A change of the sign of $\Delta_{\mathrm{odd}}$, which is linked to a changing magnitude or sign of $\Delta m_{\mathrm{I}} - \Delta m_{\mathrm{S}}$, occurs at around 100~fs, i.e. the electron thermalization time at which the stronger demagnetization at the interface due to spin transport recedes and the surface-near region starts to demagnetize due to the now dominant spin-flip scattering \cite{wieczorek2015}. \\
Note that these results demonstrate that mSHG is sensitive to spatially inhomogeneous spin dynamics on lengthscales of a few nm, which is well below the optical penetration depth in the 3$d$ transition metals of typically 10-20 nm responsible for the gradient in the initial laser excitation, and on the order of the spin-dependent electron MFPs \cite{zhukov2006, kaltenborn2014} governing spin transport. Here, mSHG can provide information about spatially inhomogeneous magnetization dynamics not available to bulk-sensitive probing via the depth sensitivity of MOKE, which is less pronounced at $d<6$~nm \cite{wieczorek2015, eschenlohr2016}. \\ 
In summary, we have demonstrated that mSHG serves as a sensitive probe for spatially inhomogeneous magnetization dynamics in epitaxial Co/Cu(001) films, due to interference of the contributions from the vacuum/Co and Co/Cu interfaces, which make up the overall mSHG signal. At $d \leq 4$~nm, it probes the sign change of the transient magnetization profile with increasing $d$ caused by the effective MFP of the fs spin current leading to demagnetization. At $d \geq 6$~nm, mSHG serves as a probe of transient magnetization profiles complementary to linear bulk-sensitive time-resolved magneto-optical measurements \cite{wieczorek2015}. This demonstrates the strength of mSHG as a probe for spatially inhomogeneous magnetization dynamics, e.g. due to spin currents in a conducting heterostructure with several interfaces.\\

\begin{acknowledgments}
	This work was funded by the German Research Foundation (DFG) through SPP 1840 QUTIF, Grant No. ES 492/1-1, and by Mercator Research Center Ruhr (MERCUR), Grant No. Pr-2014-0047. 
\end{acknowledgments}



\begin{thebibliography}{99}

\bibitem{battiato2010} M. Battiato, K. Carva, and P. M. Oppeneer, Phys. Rev. Lett. {\bf 105}, 027203 (2010).

\bibitem{melnikov2011} A. Melnikov, I. Razdolski, T. O. Wehling, E. Th. Papaiannou, V. Roddatis, P. Fumagalli, O. Aktsipetrov, A. I. Lichtenstein, and U. Bovensiepen, Phys. Rev. Lett. {\bf 107}, 076601 (2011).

\bibitem{rudolf2012} D. Rudolf {\it et al.}, Nat. Commun. {\bf 3}, 1037 (2012). 

\bibitem{malinowski2008} G. Malinowski, F. Dalla Longa, J. H. H. Rietjens, P. V. Paluskar, R. Huijink, H. J. M. Swagten, and B. Koopmans, Nat. Phys. {\bf 4}, 855-858 (2008).

\bibitem{eschenlohr2013} A. Eschenlohr, M. Battiato, P. Maldonado, N. Pontius, T. Kachel, K. Holldack, R. Mitzner, A. F\"{o}hlisch, P. M. Oppeneer, and C. Stamm, Nat. Mater. {\bf 12}, 332-336 (2013).

\bibitem{battiato2012} M. Battiato, K. Carva, and P. M. Oppeneer, Phys. Rev. B {\bf 86}, 024404 (2012).

\bibitem{schellekens2014} A. J. Schellekens, K. C. Kuiper, R. R. J. C. de Wit, and B. Koopmans, Nat. Commun. {\bf 5}, 4333 (2014).

\bibitem{belien1994} P. Beli$\ddot{\mathrm{e}}$n, R. Schad, C. D. Potter, G. Verbanck, V. V. Moshchalkov, and Y. Bruynseraede, Phys. Rev. B {\bf 50}, 9957-9962 (1994).

\bibitem{zahn1998} P. Zahn, J. Binder, I. Mertig, R. Zeller, and P. H. Dederichs, Phys. Rev. Lett. {\bf 80}, 4309-4312 (1998).

\bibitem{pan1989} R.-P. Pan, H. D. Wei, and Y. R. Shen, Phys. Rev. B \textbf{39}, 1229-1234 (1989). 

\bibitem{wieczorek2015} J. Wieczorek, A. Eschenlohr, B. Weidtmann, M. R\"osner, N. Bergeard, A. Tarasevitch, T. O. Wehling, and U. Bovensiepen, Phys. Rev. B \textbf{92}, 174410 (2015).

\bibitem{gonzalez1992} L. Gonzalez, R. Miranda, M. Salmeron, J. A. Verges, and F. Yndurain, Phys. Rev. B {\bf 24}, 3245-3254 (1981).

\bibitem{conrad2001} U. Conrad, J. G\"udde, V. J\"ahnke, and E. Matthias, Phys. Rev. B \textbf{63}, 144417 (2001).  

\bibitem{huebner1994} W. H\"ubner, K. H. Bennemann, and K. B\"ohmer, Phys. Rev. B \textbf{50}, 17597-17605 (1994). 

\bibitem{wierenga1994} H. A. Wierenga, M. W. J. Prins, D. L. Abraham, and Th. Rasing, Phys. Rev. B \textbf{50}, 1282-1285 (1994). 

\bibitem{eschenlohr2016} A. Eschenlohr, J. Wieczorek, J. Chen, B. Weidtmann, M. R\"osner, N. Bergeard, A. Tarasevitch, T. O. Wehling, and U. Bovensiepen, Proc. SPIE \textbf{9746}, Ultrafast Phenomena and Nanophotonics XX, 97461E (2016).

\bibitem{zhukov2006} V. P. Zhukov, E. V. Chulkov, and P. M. Echenique, Phys. Rev. B \textbf{73}, 125105 (2006). 

\bibitem{kaltenborn2014} S. Kaltenborn and H. C. Schneider, Phys. Rev. B \textbf{90}, 201104(R) (2014). 

\bibitem{goris2011} A. Goris, K. M. D\"obrich, I. Panzer, A. B. Schmidt, M. Donath, and M. Weinelt, Phys. Rev. Lett. {\bf 107}, 026601 (2011). 

\end{thebibliography}
\end{document}